\documentclass{article}\usepackage[]{graphicx}\usepackage[]{xcolor}
\makeatletter
\def\maxwidth{ %
  \ifdim\Gin@nat@width>\linewidth
    \linewidth
  \else
    \Gin@nat@width
  \fi
}
\makeatother

\definecolor{fgcolor}{rgb}{0.345, 0.345, 0.345}
\newcommand{\hlnum}[1]{\textcolor[rgb]{0.686,0.059,0.569}{#1}}%
\newcommand{\hlcom}[1]{\textcolor[rgb]{0.678,0.584,0.686}{\textit{#1}}}%
\newcommand{\hlopt}[1]{\textcolor[rgb]{0,0,0}{#1}}%
\newcommand{\hlstd}[1]{\textcolor[rgb]{0.345,0.345,0.345}{#1}}%
\newcommand{\hlkwa}[1]{\textcolor[rgb]{0.161,0.373,0.58}{\textbf{#1}}}%
\newcommand{\hlkwb}[1]{\textcolor[rgb]{0.69,0.353,0.396}{#1}}%
\newcommand{\hlkwc}[1]{\textcolor[rgb]{0.333,0.667,0.333}{#1}}%
\newcommand{\hlkwd}[1]{\textcolor[rgb]{0.737,0.353,0.396}{\textbf{#1}}}%

\usepackage{framed}
\makeatletter
\newenvironment{kframe}{%
 \def\at@end@of@kframe{}%
 \ifinner\ifhmode%
  \def\at@end@of@kframe{\end{minipage}}%
  \begin{minipage}{\columnwidth}%
 \fi\fi%
 \def\FrameCommand##1{\hskip\@totalleftmargin \hskip-\fboxsep
 \colorbox{shadecolor}{##1}\hskip-\fboxsep
     \hskip-\linewidth \hskip-\@totalleftmargin \hskip\columnwidth}%
 \MakeFramed {\advance\hsize-\width
   \@totalleftmargin\z@ \linewidth\hsize
   \@setminipage}}%
 {\par\unskip\endMakeFramed%
 \at@end@of@kframe}
\makeatother

\definecolor{shadecolor}{rgb}{.97, .97, .97}
\definecolor{messagecolor}{rgb}{0, 0, 0}
\definecolor{warningcolor}{rgb}{1, 0, 1}
\definecolor{errorcolor}{rgb}{1, 0, 0}
\newenvironment{knitrout}{}{} 

\usepackage{alltt}

\usepackage{natbib}
\usepackage{har2nat}
\usepackage{amsmath}
\usepackage{amssymb}

\usepackage{setspace}

\IfFileExists{upquote.sty}{\usepackage{upquote}}{}
\begin{document}

\onehalfspacing
\vspace*{0.7cm}
\begin{center}
\begin{huge}
Collected Notes on Aldrich-Mckelevey Scaling
\end{huge}
\end{center}
\vspace*{0.5cm}

\begin{large}
\begin{center}
Phil Swatton\\
University of Essex\\
p.j.swatton@essex.ac.uk
\end{center}
\end{large}

\vspace*{0.5cm}
\singlespacing
\textbf{Abstract:} Aldrich-McKelvey scaling is a method for correcting differential item functioning in ordered rating scales of perceived ideological positions in surveys. In this collection of notes, I present four findings. First, I show that similarly to ordinary least squares, Aldrich-McKelvey scaling can be improved with the use of QR decomposition during the estimation stage. While in theory this might improve accuracy, in practice the main advantage is to retain respondents otherwise lost in the estimation stage. Second, I show that this method leads to a proof of an identification constraint of Aldrich-McKelvey scaling: a minimum of three external stimuli. Third, I show that the common motivation for Aldrich-McKelvey scaling that it is robust to heteroskedasticity as compared to taking the means does not hold up. A review of the literature of prediction aggregation shows taking the mean is equally as robust. However, Aldrich-McKelvey scaling remains robust to rationalization bias and is transparent in its assumptions. Finally, I show that the setup of Bayesian Aldrich-McKelvey Scaling and Aldrich-McKelvey scaling differ from each other in their parameterisation. This is not commonly acknowledged in the literature, and new users of these methods should be aware.
\vspace*{1cm}

\thispagestyle{empty}

\newpage
\doublespacing
\setcounter{page}{1}


\section{Introduction}

Aldrich-McKelvey scaling is a method designed to correct differential item functioning in survey respondent placements of external stimuli along ordered rating scales, such as the ideological positions of political parties \cite{aldrich1977}. In doing so, it allows for recovery of corrected stimuli positions and by extension respondent positions or ideal points along the same scales. Differential item functioning can be understood as survey respondents perceiving or understanding the scale in fundamentally different ways - thus leading to differing placements of the stimuli. Aldrich-McKelvey scaling has sometimes also been discussed in the context of rationalisation bias, where respondents have a tendency to shift the placements of parties they dislike away from themselves and the placements of parties they like towards themselves. However, it is only in the form of a much more recent adaptation that rationalisation bias has been explicitly included in the model, although Aldrich-McKelvey scaling remains robust in the case of estimating stimuli positions and reasonably robust in the case of estimating respondent positions \cite{bolstad2020}.

This more recent adaptation was itself based on an adaptation of the Aldrich-Mckelvey scaling method in the form of Bayesian Aldrich-McKelvey scaling \cite{hare2015}. This has also been developed in an alternative direction, with Ordered Bayesian Aldrich-McKelvey scaling being first proposed at the 2019 MPSA annual meeting \cite{cikanek2019}. In a different, older vein, the method was also generalised to a multidimensional case - albeit one in which the ability to scale respondents and stimuli simultaneously was lost \cite{poole1998}. Although not well-known outside political science (and perhaps not even within), application-specific methods such as Aldrich-McKelvey scaling represented in the estimation of Gary King the most satisfactory approach to correction for differential item functioning \citeyear{king2004}. The method is unsurprisingly particularly popular in work exploring the spatial model of political competition. Results from Aldrich-McKelvey scaling have been used to generate estimates of voter information \cite{palfrey1987}, in scaling elite positions \cite{saiegh2009}, in testing Downsian theories of voter preference \cite{hollibaugh2013}, in placing voters and parties from different countries on the same scale \cite{lo2014}, and in assessing the role of valence in voter utility functions \cite{gouret2021}.

What makes Aldrich-Mckelvey scaling and its derivatives unique as compared to other more generic methods such as factor analysis, item response theory, or multidimensional scaling is the fact that it scales the stimuli at \textit{the same level of input} in which they exist. In other words, where other scaling methods necessarily reduce the dimensionality of the data they take as input, Aldrich-Mckelvey scaling retains the same dimensionality for both stimuli and respondents. What it instead does is assign variation in stimuli placements to a set of parameters broadly capturing the process of differential item functioning. This is not to say that models allowing the recovery of higher dimensions are not useful or theoretically important. Rather, this unique feature of Aldrich-McKelevey scaling is useful both for practical purposes and for better exploration of respondent variation in lower dimensionalities. While not well-known outside the world of political science, this may in fact give rise to more use in other social sciences, such as psychology or sociology, where the placement of external stimuli along some continuum may in fact be useful at a particular level of analysis.

Despite the method's usefulness, its several adaptations, and its popularity in certain parts of the political science community, it remains surprisingly under-explored. Widely popular methods such as ordinary least squares are well-understood, with several important results and findings regarding best practice for estimation, identification conditions, and performance under the violation of assumptions. My purpose in this collection of notes is to present several small, key results in these areas, with further comment on differences between Aldrich-Mckelvey scaling and Bayesian Aldrich-Mckelvey scaling. The first note discusses the possibility of estimating the Aldrich-Mckelvey solution through QR decomposition, thus allowing for greater numerical accuracy and better respondent retention. This estimation strategy in turn gives rise to the second note, which shows that the Aldrich-Mckelvey model is not identified when the number of stimuli is below 3. The third note challenges a key assumption in the Aldrich-Mckelvey literature, which is that it is especially robust to heteroskedasticity as compared to mean-centering. As we know from the forecasting aggregation literature, taking the mean is also robust to heteroskedasticity. There remain however good reasons for favouring the Aldrich-McKelvey scaling method seperate from performance assessments. The fourth and final note discusses the fact that the Bayesian Aldrich-Mckelvey parameters as set up in the model necessarily have a different interpretation to the original Aldrich-Mckelvey model, setting asside conceptual differences in the estimation approaches.

\section{Aldrich-Mckelvey Scaling}

Before proceeding to the first note, I provide a brief sketch of Aldrich-McKelvey scaling and its traditional estimation via a least-squares solution. As above Aldrich-McKelvey scaling is a method for correcting differential item functioning in the placement of external stimuli along an ordered rating scale. The method does not \textit{require} that the data be ordinal (and in fact treats them as continuous), but in practice the data available to researchers are always ordinal. Let $n$ be the number of respondents indexed by $i$ and $J$ be the number of external stimuli indexed by $j$. We then place the $J$ placements of stimuli by respondent $i$ in a matrix with a column of 1s:
\begin{equation}
  \boldsymbol{X}_{i} = \begin{bmatrix} 1 & \boldsymbol{X}_{i1} \\ 1 & \boldsymbol{X}_{i2} \\ \vdots & \vdots \\ 1 & \boldsymbol{X}_{iJ} \end{bmatrix}
  \label{eq:Xi}
\end{equation}

Aldrich-McKelvey write a model of the form
\begin{equation}
  c_i + w_iX_{ij} = Y_j + u_{ij}
  \label{eq:ammodel}
\end{equation}
where $c_i$ is the intercept or shift parameter for respondent $i$ and $w_i$ is the weight or scaling parameter. I retain the use of 'intercept' and 'weight' respectively in the remainder of this paper. In the Aldrich-McKelvey model, these parameters govern the differential item functioning process. $\boldsymbol{Y_j}$ represents the true position of the $j$th stimulus, while $u_{ij}$ is an error term, for which Gauss-Markov assumptions apply. To obtain the solution to the above model, set:
\begin{equation}
  \boldsymbol{A} = \sum_{i=1}^n{\boldsymbol{X}_{i}'(\boldsymbol{X}_{i}'\boldsymbol{X}_{i})^{-1}\boldsymbol{X}_{i}'}
  \label{eq:A}
\end{equation}

Aldrich and McKelvey prove that the that the Aldrich-McKelvey estimator $\hat{\boldsymbol{Y}}$ of the `true' stimuli positions $\boldsymbol{Y}$ is the eigenvector of
\begin{equation}
  \boldsymbol{A} - n\boldsymbol{I}
  \label{eq:amsolution}
\end{equation}
corresponding to the highest negative eigenvalue of the above (i.e. the negative eigenvalue closest to 0). With the Aldrich-McKelvey scaling method briefly established \cite[see][for the full proof]{aldrich1977}, I now turn to the four notes that compose this paper.

\section{Estimation by QR Decomposition}

A fact well-known among statistical software developers but much less well-known in the wider social science community is that the ordinary least squares model (OLS) is not estimated through the well-known matrix formula $\boldsymbol{\hat{\beta}} = (\boldsymbol{X}'\boldsymbol{X})\boldsymbol{X}'\boldsymbol{Y}$. Instead, OLS is typically estimated through some form of \textit{matrix decomposition} - usually \textit{QR decomposition}. This is due to the fact that floating-point computations contain some inherent inaccuracy \cite[see][]{goldberg1991}, thus in turn rendering matrix inversion to be particularly numerically unstable. \textsf{R}'s \textsf{lm} function in particular represents a popularly-used function that estimates OLS parameters through QR decomposition. It follows that Aldrich-McKelvey scaling may also benefit from an alternative approach to estimation. In this note, I show how QR decomposition can also be used for the purpose of estimating the Aldrich-McKelvey scaling results.

First, a brief examination of \eqref{eq:A} shows that Aldrich-McKelvey scaling relies on the calculation of
\begin{equation}
  \boldsymbol{A}_i = \boldsymbol{X}_{i}(\boldsymbol{X}_{i}'\boldsymbol{X}_{i})^{-1}\boldsymbol{X}_{i}'
  \label{eq:Ai}
\end{equation}
for all $n$ units. In practice, for some respondents $(\boldsymbol{X}_{i}'\boldsymbol{X}_{i})^{-1}$ may not be invertible. When this is the case, the information provided by some respondents is lost and we use $<n$ respondents in calculating the stimuli positions. However, by using QR decomposition, it becomes possible not only to more accurately calculate $\boldsymbol{A}_i$, but also to retain a greater number of respondents in the overall estimation. While atypical in presenting the QR decomposition, I keep the $i$ index throughout to better relate the decomposition to calculating \eqref{eq:Ai} for each respondent. The QR decomposition of a matrix is given by
\begin{equation}
  \boldsymbol{X}_i = \boldsymbol{Q}_i\boldsymbol{R}_i
  \label{eq:QR}
\end{equation}
where $\boldsymbol{R}_i$ is an upper triangular matrix and $\boldsymbol{Q}_i$ is of the same dimensions of $\boldsymbol{X}_i$ and its columns are orthonormal, meaning that
\begin{equation}
  \boldsymbol{Q}_i'\boldsymbol{Q}_i = \boldsymbol{I}
  \label{eq:QQ}
\end{equation}

Substituting the QR decomposition from \eqref{eq:QR} into the \eqref{eq:Ai} we obtain
\begin{equation}
  \boldsymbol{A}_i = \boldsymbol{Q}_i\boldsymbol{R}_i(\boldsymbol{R}_i'\boldsymbol{Q}_i'\boldsymbol{Q}_i\boldsymbol{R}_i)^{-1}\boldsymbol{R}_i'\boldsymbol{Q}_i'
  \label{eq:aqrp1}
\end{equation}
recalling \eqref{eq:QQ} this simplifies to :
\begin{equation}
  \boldsymbol{A}_i = \boldsymbol{Q}_i\boldsymbol{R}_i(\boldsymbol{R}_i'\boldsymbol{R}_i)^{-1}\boldsymbol{R}_i'\boldsymbol{Q}_i'
  \label{eq:aqrp2}
\end{equation}

The rules of algebraically manipulating matrices dictate that
\begin{equation}
  (\boldsymbol{R}_i'\boldsymbol{R}_i)^{-1} = \boldsymbol{R}_i^{-1}(\boldsymbol{R}_i')^{-1}
  \label{eq:rr}
\end{equation}
and we can thus substitute \eqref{eq:rr} into \eqref{eq:aqrp2} to obtain:
\begin{equation}
  \boldsymbol{A}_i = \boldsymbol{Q}_i\boldsymbol{R}_i\boldsymbol{R}_i^{-1}(\boldsymbol{R}_i')^{-1}\boldsymbol{R}_i'\boldsymbol{Q}_i
  \label{eq:aqrp3}
\end{equation}
which by the definition of an inverse and an identity matrix simplifies to
\begin{equation}
  \boldsymbol{A}_i = \boldsymbol{Q}_i\boldsymbol{Q}_i'
  \label{eq:aqrp4}
\end{equation}
thus drastically simplifying \eqref{eq:A}. This simplification avoids the need to invert
\begin{equation}
  \boldsymbol{X}_{i}'\boldsymbol{X}_{i}
  \label{eq:xx}
\end{equation}
which as above not only is numerically unstable in practice but also results in the unnecessary loss of some respondents from the calculation of placements for the stimuli. This solution instead reduces the problem of estimating \eqref{eq:Ai} to a problem of estimating the QR decomposition for each respondent, rather than performing a direct inversion for each respondent. However, this result for the purpose of estimation also aids us in better understanding the identification conditions for Aldrich-McKelvey scaling.

\section{Identification}

An important consideration for many types of latent variable model or scaling method is that to some degree model identification is driven by the need to put information in to get information out. Factor analysis - both exploratory and confirmatory - has strict rules on the number of parameters, variances, and covariances being estimated relative to the number of variances and covariances being inputed. Principle components analysis estimates only as many principle components as input vectors. Similarly, to estimate parameters in a linear model we need enough input data.
The question then follows - what are the identification constraints for Aldrich-McKelvey scaling? While Aldrich and McKelvey place some scale constraints (e.g. mean 0) \cite{aldrich1977}, this is to prevent indeterminacy in the solution for any value of $J$. The question here is what is the minimum $J$ for the model to be identified - i.e. for a unique solution to exist for some given data. Using QR decomposition, it is possible to show that no solution exists once $J < 2$. The result that follows begins by proving this in the case of $J = 2$, before discussing the case of $J = 1$.

\subsection{No Unique Solution when $J=2$}

We begin by recalling the QR-decomposition based approach to calculating $\boldsymbol{A}$ for the Aldrich-McKelvey solution:
Recall from the new approach to estimating the Aldrich-McKelvey solution via the QR decomposition that
\begin{equation}
  \boldsymbol{A} = \sum_{i=1}^n\boldsymbol{A}_i = \sum_{i=1}^n\boldsymbol{Q}_i\boldsymbol{Q}_i'
  \label{eq:aiSum}
\end{equation}

When $J = 2$, each $\boldsymbol{X}_i$ matrix and thus each $\boldsymbol{Q}_i$ matrix will be square. When the QR decomposition is performed on a square matrix, the Q matrix will be an orthogonal matrix meaning that
\begin{equation}
  \boldsymbol{Q}_i'\boldsymbol{Q}_i = \boldsymbol{Q}_i\boldsymbol{Q}_i' = \boldsymbol{I} \Rightarrow \boldsymbol{Q}_i' = \boldsymbol{Q}_i^{-1}
  \label{eq:squareQR}
\end{equation}
it follows that in this special case:
\begin{equation}
  \boldsymbol{A} = \sum_{i=1}^n\boldsymbol{Q}_i\boldsymbol{Q}_i' = \sum_{i=1}^n\boldsymbol{I} = n\boldsymbol{I}
  \label{eq:squareA}
\end{equation}

If we insert \eqref{eq:squareA} into \eqref{eq:amsolution} we obtain
\begin{equation}
  n\boldsymbol{I} - n\boldsymbol{I}
  \label{eq:squareAM}
\end{equation}
meaning that when $J=2$ the eigendecomposition will be performed on a $2 \times 2$ $\boldsymbol{0}$ matrix.

When an eigendecomposition is performed on any $\boldsymbol{0}$ matrix, all eigenvalues will be 0 and thus the number of corresponding eigenvectors of the correct proportion will be infinite. In other words, there is no solution. This follows first because there is no negative eigenvalue (recall that the Aldrich-McKelvey solution selects based on the highest negative eigenvalue); and second even ignoring this the range of corresponding eigenvectors to a 0 eigenvalue is infinte. There is therefore no unique solution, and thus the scaling method is not identified when $J=2$. Researchers should be cautious in this case: the fact that the result has an infinite range of solutions does not mean that software will not present a solution. Running base \textsf{R}'s \textsf{eigen} function on a $2 \times 2$ $\boldsymbol{0}$ matrix produces a matrix with two eigenvectors containing the arbitrary values of -1 and 1:
\begin{equation}
\begin{bmatrix} 0 & -1 \\ 1 & 0\end{bmatrix}
\label{eq:basicspaceSquareAM}
\end{equation}

Similarly and more importantly, the \textsf{aldmck} function from the \textsf{basicspace} \textsf{R} package will produce the values of 0 and 1 for its stimuli placement estimates despite correctly returning a full set of 0s for the extracted eigenvalues \cite{poole2016}. Researchers utilising this package (and given it is to my knowledge the only package making the Aldrich-McKelvey scaling method widely accessible they are likely to be) should take care in interpreting its results - any rescaling performed when $J<3$ will necessarily be meaningless as they stem from a solution which in practice is not identified and thus the results themselves could change arbitrarily.

\subsection{No Unique Solution when $J=1$}

Until this point, in this note I have focussed on the case where $J=2$ while making claims regarding the broader case of $J<3$. I now turn to the case of $J=1$. While intuitively it makes sense for a model to be unidentified when $J=1$ if it is unidentified when $J=2$, it is nonetheless worth establishing this beyond the realm of intuition. First, estimation of \eqref{eq:aqrp4} cannot occur when $J=1$ because the QR decomposition only exists when the number of rows is greater than or equal to the number of columns. It follows that the calculating $\boldsymbol{A}$ must necessarily fall back to \eqref{eq:A} (insofar as other methods of calculating $\boldsymbol{A}$ have not yet been discovered.

However, in the case of $J=1$ calculation of each $\boldsymbol{X}_i'\boldsymbol{X}_i$ takes on a particular behaviour that prevents its own inversion. To show this, it is worth demonstrating how $\boldsymbol{X}_i'\boldsymbol{X}_i$ is calculated if its two component values are not known when we have two columns and one row. We begin by setting the two values of $\boldsymbol{X}_i$ to $a$ and $b$:
\begin{equation}
  \boldsymbol{X}_i = \begin{bmatrix} a & b \end{bmatrix}
  \label{eq:abmat}
\end{equation}

Taking the inner product of \eqref{eq:abmat} will take the form
\begin{equation}
  \boldsymbol{X}_i'\boldsymbol{X}_i = \begin{bmatrix} a \\ b \end{bmatrix}\begin{bmatrix} a & b \end{bmatrix}
\end{equation}
which necessitates the following multiplications of $a$ and $b$:
\begin{equation}
  \boldsymbol{X}_i'\boldsymbol{X}_i = \begin{bmatrix} a^2 & b*a \\ a*b & b^2 \end{bmatrix}
  \label{eq:abmutiplication}
\end{equation}

In the case of Aldrich-McKelvey scaling, by the defintion of $\boldsymbol{X}_i$ in \eqref{eq:Xi} we know that $a=1$. Inserting this into \eqref{eq:abmutiplication} yields the following solution for $\boldsymbol{X}_i'\boldsymbol{X}_i$:
\begin{equation}
  \boldsymbol{X}_i'\boldsymbol{X}_i = \begin{bmatrix} 1 & b \\ b & b^2 \end{bmatrix}
\end{equation}

A clear feature of the $J=1$ solution for $\boldsymbol{X}_i'\boldsymbol{X}_i$ is that the second row is a simply the first multiplied by $b$ (and similarly, the second column is simply the first multiplied by $b$). This means that the matrix $\boldsymbol{X}_i'\boldsymbol{X}_i$ is singular and its inverse does not exist. While it is therefore possible to show that a solution does not exist for the Aldrich-McKelvey solution does not exist either via traditional estimation or via the QR decomposition above due to the impossibility of calculating $\boldsymbol{A}$, a more conclusive proof of the type existing for the case of $J=2$ does not yet exist. It is likely that other matrix decompositions could be used to calculate $\boldsymbol{A}$ - we must rely on the non-existence of a solution when $J=2$ and a the non-existence of a solution for both approaches to estimation when $J=1$ to inductively conclude that a solution does not exist when $J=1$.

\section{Central Tendency as an Estimator of Stimulus Position}

It is frequently asserted that the advantage of Aldrich-McKelvey scaling relative to simply taking the mean value of a vector of placements is that Aldrich-McKelvey scaling is robust to heteroskedasticity in the error term $u_{ij}$ in \eqref{eq:ammodel}. This claim originates from a paper by Palfrey and Poole \citeyear{palfrey1987}, where they show that Aldrich-McKelvey scaling is robust to heteroskedasticity in the $u_{ij}$ in recovering stimuli placements but has a bias towards unidimensional results that becomes stronger for low-information respondents. After this initial piece of research, the claim that Aldrich-McKelvey scaling should be preferred to taking the mean \textit{because it is robust to heteroskedasticity where taking the mean is not} proliferated \cite[see e.g.][]{armstrong2020, gouret2021}. However, by drawing comparisons with findings from the world of forecasting and forecasting aggregation, it becomes possible to realise that taking the mean is just as robust to heteroskedasticity as Aldrich-McKelvey scaling.

The comparability of forecast aggregation and Aldrich-McKelvey scaling stems from the fact that both represent `crowd wisdom' approaches to estimating some quantity of interest. In the case of forecast aggregation, our interest is in attempting to quantify the probability of some future event occurring. In the case of Aldrich-McKelvey scaling, our interest is instead in translating the `common sense' of voters regarding party positions into quantifiable party placements alongside respondent placements. The difference of course is that in the case of forecast aggregation, we eventually learn whether an event occurs or not and in the long term can learn how accurate different aggregators are. By contrast, with Aldrich-McKelvey scaling and its derivatives we never learn at any point in time what the `true' position of some political party is along some dimension of interest. We are therefore constrained to relying on a combination of assumptions about the data generating process, things we can actually learn about the data generating process without observing `true' values, and simulations when seeking to understand the behaviour of the Aldrich-McKelvey scaling method.

Of interest to me in this note is the `Bias-Information-Noise' (BIN) model for assessing forecasts \cite{satopaa2021}. As suggested by the name, this model decomposes the results of forecasts into three key components. Importantly, this allows us to assess how different forecast aggregators actually work - do they reduce bias, increase the information recovered, or simply reduce noise? If $\theta$ is some estimated value and $T$ is the `true' value, then the BIN model can be summarised in the following three formulae:

\begin{gather}
  Bias = E[\theta - T]\index{eq:bias} \nonumber\\
  Information = cov(\theta,T)\index{eq:info} \label{eq:bin}\\
  Noise = var(\theta) - var(T)\index{eq:noise} \nonumber
\end{gather}

An important finding from the world of forecast aggregation is that taking the mean is an effective way of reducing \textit{noise}, but not necessarily of increasing \textit{information} or reducing \textit{bias} \cite{satopaa2017, satopaa2021}. The implication here is that the mean will reduce any form of \textit{noise} but will not remove structural biases. Whether that noise increases or decreases linearly with some variable - such as the amount of information a respondent possesses - is not necessarily important. Taking the mean will still reduce the noise to a single central point. Given that Palfrey and Poole \citeyear{palfrey1987} only assess the violation of the homoskedasticity Gauss-Markov assumption and thus is not a form of \textit{bias} per se, there is no reason to assume that taking the mean should be any less robust to heteroskedasticity in \eqref{eq:ammodel}. Indeed, this is directly testable. Taking the simulation code demonstrating the robustness of Aldrich-McKelvey scaling in the article for the \textsf{basicspace} \textsf{R} package \cite{poole2016}, it is easy to see that the means are infact highly correlated with the Aldrich-McKelvey solution:

\begin{knitrout}
\definecolor{shadecolor}{rgb}{0.969, 0.969, 0.969}\color{fgcolor}\begin{kframe}
\begin{alltt}
\hlcom{# For replicability}
\hlkwd{set.seed}\hlstd{(}\hlnum{1234}\hlstd{)}

\hlcom{# Stimuli positions}
\hlstd{J} \hlkwb{<-} \hlnum{6}
\hlstd{Yj} \hlkwb{<-} \hlkwd{rnorm}\hlstd{(}\hlnum{6}\hlstd{,} \hlkwc{mean}\hlstd{=}\hlnum{0}\hlstd{,} \hlkwc{sd}\hlstd{=}\hlnum{1}\hlstd{)}
\hlstd{Yj} \hlkwb{<-} \hlstd{(Yj}\hlopt{-}\hlkwd{mean}\hlstd{(Yj))}\hlopt{/}\hlkwd{sd}\hlstd{(Yj)}

\hlcom{# Respondent error}
\hlstd{N} \hlkwb{<-} \hlnum{500}
\hlstd{resp_sd} \hlkwb{<-} \hlkwd{runif}\hlstd{(N,} \hlkwc{min}\hlstd{=}\hlnum{0.3}\hlstd{,} \hlkwc{max}\hlstd{=}\hlnum{0.9}\hlstd{)}
\hlstd{ui} \hlkwb{<-} \hlkwd{matrix}\hlstd{(}\hlnum{NA}\hlstd{, N, J)}
\hlkwa{for} \hlstd{(i} \hlkwa{in} \hlnum{1}\hlopt{:}\hlstd{N) ui} \hlkwb{<-} \hlkwd{rnorm}\hlstd{(N,} \hlkwc{mean}\hlstd{=}\hlnum{0}\hlstd{,} \hlkwc{sd}\hlstd{=resp_sd)}

\hlcom{# Respondent intercepts and weights}
\hlstd{wi} \hlkwb{<-} \hlkwd{runif}\hlstd{(N,} \hlkwc{min}\hlstd{=}\hlnum{0}\hlstd{,} \hlkwc{max}\hlstd{=}\hlnum{1}\hlstd{)}
\hlstd{ci} \hlkwb{<-} \hlkwd{rnorm}\hlstd{(N,} \hlkwc{mean}\hlstd{=}\hlnum{0}\hlstd{,} \hlkwc{sd}\hlstd{=}\hlnum{1}\hlstd{)}

\hlcom{# Respondent placements}
\hlstd{Yij} \hlkwb{<-} \hlstd{(}\hlkwd{rep}\hlstd{(}\hlnum{1}\hlstd{,N)} \hlopt{%o%} \hlstd{Yj)} \hlopt{+} \hlstd{ui}
\hlstd{Xij} \hlkwb{<-} \hlstd{(}\hlnum{1}\hlopt{/}\hlstd{wi)} \hlopt{*} \hlstd{(Yij} \hlopt{-} \hlstd{ci)}

\hlcom{# aldmck from basicspace}
\hlstd{am} \hlkwb{<-} \hlkwd{aldmck}\hlstd{(Xij,} \hlkwc{polarity}\hlstd{=}\hlnum{1}\hlstd{)}
\hlstd{stim} \hlkwb{<-} \hlstd{am}\hlopt{$}\hlstd{stimuli}

\hlcom{# Taking the means}
\hlstd{means} \hlkwb{<-} \hlkwd{apply}\hlstd{(Xij,} \hlkwc{MARGIN}\hlstd{=}\hlnum{2}\hlstd{,} \hlkwc{FUN}\hlstd{=mean)}
\end{alltt}
\end{kframe}
\end{knitrout}

The above code is adapted from Poole et. al. \citeyear{poole2016}, with changes to the code to make it consistent with the notation I've used in these notes and the addition of a seed to facilitate replicability. Note the heteroskedastic error $u_i$ instead of the previous error term $u_{ij}$ from \eqref{eq:ammodel}. I have also slightly tidied some steps to simplify the code. In line with past results, the stimuli placements are closely matched to the `true' stimuli positions with a Pearson's correlation of 1. However, so too are the means with a Pearson's correlation of 1 (both are not exactly 1 but have been rounded to the nearest 3 decimal places, giving a correlation of effectively 1). Indeed, the correlation between the recovered stimuli positions from the Aldrich-McKelvey scaling and simply taking the means is 1 (once again after rounding). Given the similarity between these results and the relative ease of taking the means, why then use Aldrich-McKelvey scaling?

The first answer to the above questions is that it renders the assumptions of the measurement process \textit{transparent}. When we perform Aldrich-McKelvey scaling, there are clear assumptions regarding the data generating process as articulated in the original paper \cite{aldrich1977}. It is clear that insofar as recovery of the stimuli is concerned, both methods are robust to heteroskedasticity in \eqref{eq:ammodel}. But when we take the mean, it is no longer clear whether we are assuming a data generating process of the form in \eqref{eq:ammodel} with all the corresponding assumptions. Similarly, while we \textit{could} regress the mean placements on the respondent placements to obtain similar parameters, there is less transparency in the process by which we obtain these parameters and the assumptions we make in the process. Where the assumptions of the Aldrich-McKelvey scaling method \textit{are} violated, this creates space for the development of new scaling methods that can be built on this initial attempt.

Secondly, the Aldrich-McKelvey scaling process is robust not merely to heteroskedasticity, but also to rationalisation bias. Simulations in \citeauthor{bolstad2020} \citeyear{bolstad2020} shows that Aldrich-McKelvey scaling still reproduces accurate estimates of stimuli placements under the presence of rationalisation bias, and remains somewhat accurate in its estimates of respondent placements on the same scale. There is no reason to imagine that merely taking the mean will be equally robust. We should thus focus on this advantage of Aldrich-McKelvey scaling in motivating its use, rather than on its robustness to heteroskedasticity relative to taking the means of the scales. However, the issue of how both methods respond to extreme rationalisation bias in practice remains to be further explored.

\section{BAM is not AM}

By far one of the most popular adaptations of Aldrich-McKelvey scaling is Bayesian Aldrich-McKelvey scaling. The popularity of the Bayesian variant is unsurprising given two key limitations of the original version. First, the only process developed for obtaining error bounds on the Aldrich-McKelvey stimuli placements is bootstrapping. Second and more importantly, Aldrich-McKelvey scaling has strict requirements for the number of placements respondents make - if some respondent places say 4 out of 6 stimuli, the information that respondent possesses is necessarily lost. While the `blackbox' scaling method allowed the presence of missing data \cite{poole1998}, this method necessarily scales data to a higher dimensionality and does not scale respondents and stimuli simultaneously. Bayesian Aldrich-McKelvey scaling fixes both of these issues - first by allowing missing data in stimuli placements; and second by using the posterior distributions for

These gains facilitate a number of research applications not previously imaginable for Aldrich-McKelvey scaling. First, where the anchoring vignettes first proposed to serve as a means of establishing the comparability of different stimuli placements across space and time \cite{king2004} had previously been used in the `blackbox' scaling method \cite{bakker2014}; Bayesian Aldrich-McKelvey scaling now allowed this to be performed at the lower level of dimensionality \cite{jessee2021}. This allows not only for stimuli positions to be compared at lower dimensionalities over time and space, but also for respondent positions along the same scale. None of this necessarily precludes exploring the higher dimensionalities or `basic space' of ideology - but it does allow these explorations to begin from scales corrected for differential item functioning.

However, a point that has not yet appeared in the literature is the fact that Bayesian Aldrich-McKelvey scaling is built on a slightly different data generating process to the original Aldrich-McKelvey scaling model. Recall that the Aldrich-McKelvey model was of the form
\begin{equation}
  c_i + w_iX_{ij} = Y_j + u_{ij} \tag{\ref{eq:ammodel} repeated}
\end{equation}

By contrast, the Bayesian-Aldrich McKelvey scaling model is of the form \cite{hare2015}
\begin{equation}
  a_i + b_iY_j = X_{ij} + u_{ij}
  \label{eq:bammodel}
\end{equation}
where $a_i$ is the intercept parameter and $b_i$ is the weight parameter. The notable difference from \eqref{eq:ammodel} to \eqref{eq:bammodel} is the fact the parameters operate on $Y_j$ instead of $X_{ij}$. Insofar as I have been able to tell, this is not a deliberate change - the same difference is repeated in describing the original Adlrich-McKelvey model in the \textit{Analysing Spatial Models of Judgement of Choice} book \cite{armstrong2020}. This is not necessarily straightforwardly important, but it does mean the parameters have differing interpretations. Arguably, the interpretation of the Bayesian Aldrich-McKelvey parameters is more in line with with our interpretation of the model in that they `shift' and `stretch' the true stimulus position to the perceived one thus directly capturing the differential item functioning process. By contrast, the Aldrich-McKelvey parameters relate the perceived stimulus to the `true' stimulus. Insofar as the difference is a mistake it is therefore a happy one. This is worth noting not just for comparisons between the individual-level parameters of the results.
This difference between the models has also made it into adaptations of the Bayesian Aldrich-McKelvey scaling method such as Ordered Bayesian Aldrich-McKelvey scaling \cite{cikanek2019} and the Intercept-Stretch-Rationalization model \cite{bolstad2020}.

\section{Conclusion}

Aldrich-McKelvey scaling and its adaptations remain under-utilised within political science. They represent some of our best methods for placing both external stimuli such as political parties and leaders and survey respondents on the same scale while corrected for differential item functioning and rationalization bias. With the advent of the blackbox scaling method and the later creation of Bayesian Aldrich-McKelvey scaling, these methods can be combined with anchoring vignettes to achieve comparable position placements over time and space. The potential of these scaling methods does not necessarily need to be constrained to political science. Uniquely among latent variable models and scaling methods, Aldrich-McKelvey scaling retains the same lower dimension in its output, where other scaling methods recover a higher dimensionality. There is little reason to imagine that Aldrich-McKelvey scaling could not be used to scale stimuli in other fields - particularly where respondents can be placed along the same scale.

In this collection of notes, I have sought to improve the understanding of the scaling method. In the first note, I show how QR decomposition can be used to produce more accurate results and retain a larger number of respondents. In the second note, I show that the model is not identified when the number of stimuli is below 3. In the third note, I show how the heteroskedasticity-robustness of the method is not the reason to favour it over taking the mean. Instead, the strong point of the Aldrich-McKelvey method is the fact it makes our assumptions regarding the data generating process transparent and in the longer term facilitates new adaptations for situations where these assumptions are not met. Finally, I show that the Aldrich-McKelvey scaling method is different in form to the Bayesian Aldrich-McKelvey scaling method and thus in interpretation of the individual parameters.

\bibliography{00_bibliography}

\end{document}